\documentclass{aa}
\usepackage{graphicx}
\begin{document}
   \title{Pulsar's kicks and $\gamma$-ray bursts}

 %  \subtitle{I. Overviewing the $\kappa$-mechanism}

   \author{X. H. Cui,
         \inst{1}
          H. G. Wang,
          \inst{2}
          R. X. Xu,
          \inst{1}
          \and
          G. J. Qiao \inst{1}%\fnmsep\thanks{Just to show the usage
%          of the elements in the author field}
          }

   \offprints{R. X. Xu (r.x.xu@pku.edu.cn)}

   \institute{Astronomy Department, School of Physics, Peking University, Beijing 100871, China\\
             \email{(xhcui,rxxu,gjq)@bac.pku.edu.cn}
         \and
             Center for Astrophysics, Guangzhou University, Guangzhou 510400, China\\
             \email{cosmic008@263.net}
%                     accept e-mails}
             }

  % \date{Received September 15, 1996; accepted March 16, 1997}
   \date{Received~~~~~~~~~~~~~~~~~; ~~accepted}

% \abstract{}{}{}{}{}
% 5 {} token are mandatory

  \abstract
{}
{The consistence of the distributions of pulsar's kick
velocities from the model of GRB and from the pulsar observations
is tested based on the supernova-GRB ($\gamma$-ray burst)
association and under the assumption that the GRB asymmetric
explosions produce pulsars.}
{The deduced distribution of kick velocity from the model of GRB
and the observed kick distribution of radio pulsars are checked by
K-S test.}
{These two distributions are found to come from a same parent
population.}
{This result may indicate that GRBs could be really related to
supernova, and that the asymmetry of GRB associated with supernova
would cause the kick of pulsars.}


   \maketitle
%
%________________________________________________________________

\section{Introduction}

The difficulty of reproducing two kinds of astronomical bursts are
challenging today's astrophysicists to find realistic explosive
mechanisms.
On one hand, $\gamma$-ray bursts (GRBs) are puzzling phenomena,
the center engine of which is still an outstanding problem
although the related fireball models have been well-developed due
to accumulation of more and more observational data. The launch of
Swift is stimulating the study in this area (see Zhang 2007 for a
recent review).
On the other hand, the failure to simulate supernovae (SNe)
successfully in the neutrino-driven explosion model troubles
astrophysicists over long time, and the call for alternative
mechanisms grew stronger and stronger (Mezzacappa 2005; Buras et al.
2003).

The discovery of 4 clear associations (and many SN bumps in the late
optical afterglow light curves) between long soft GRBs and Type
II/Ibc SNe (see, e.g., the review by Woosley \& Bloom 2006) results
in finding common explosive processes for SNe and GRBs: to form
spinning rapidly black holes (Woosley 1993), neutron stars
(Klu\'zniak \& Ruderman 1998), or even quark stars (Dai \& Lu 1998).
It is worth noting that GRB as a signature of phase transition to
quark-gluon plasma (Xu et al. 1999; Wang et al. 2000; Yasutake et
al. 2005; Paczy\'nski \& Haensel 2005; Drago, Pagliara \& Parenti
2007; Haensel \& Zdunik 2007) has also great implications in the
study of elementary interactions between quarks.
It was addressed that the {\em bare} quark surfaces could be
essential to successful explosions of both GRBs and SNe (Xu 2005;
Paczy\'nski \& Haensel 2005; Chen \& Xu 2006) because of the
chromatic confinement (the photon luminosity of a quark surface is
not then limited by the Eddington limit).
For simplicity, 1-dimensional (i.e., spherically symmetric)
calculation of Chen \& Xu (2006) shows that the lepton-dominated
fireball supported by a bare quark surface do play a significant
role in the explosion dynamics under such a photon-driven
scenario.
However, what if the expanding of a fireball outside quark surface
is not spherically symmetric?
That asymmetry may naturally result in kicks of quark stars. But
how to test this idea? These issues will be focused here.

Quark stars could well reproduce the observational features of
pulsar-like stars (Xu 2006).
Radio pulsars have long been recognized to have great space
velocities (e.g. Gunn \& Ostriker 1970; Lorimer, Bailes, \&
Harrison 1997; Lyne, Anderson \& Salter 1982; Cordes \& Cherno
1998). Lyne \& Lorimer (1994) have observed a large mean velocity
of pulsars $\upsilon \approx 450\pm 90$ km s$^{-1}$. From a
comparison with Monte Carlo simulation, Hansen \& Phinney (1997)
found that the mean birth speed of a pulsar is $\sim 250-300$ km
s$^{-1}$. Applying the recent electron density model to determine
pulsar distances, Hobbs et al. (2005) gave mean two-dimensional
(2D) speeds of $246\pm 22$ km s$^{-1}$ and $54\pm 6$ km s$^{-1}$
for the normal and recycled pulsars, respectively.

However, the origin of these kicks is still a matter of debate. In
1994, Lyne \& Lorimer suggested generally that any small asymmetry
during the explosion can result in a substantial ``kick'' to the
center star. From the observation of binary pulsar, Lai, Bildsten
\& Kaspi (1995; see also Lai 1996) proposed also that the pulsar
acquired its velocity from an asymmetric SN collapse. In 1996,
Burrows \& Hayes argued that an anisotropic stellar collapse could
be responsible for pulsar kicks. Cen (1998) proposed that a SN
produces a GRB and a strong magnetized, rapidly rotating NS
emitting radio pulse, and, for the first time, the author related
the kicks of pulsars to GRBs. Dar \& Plaga (1999) proposed that
the natal kick may arise from the emission of a relativistic jet
of its center compact star. Lai, Chernoff \& Cordes (2001) pointed
out three kick mechanisms: electromagnetic rocket mechanism
(Harrison \& Tademaru 1975), hydrodynamically driven and
neutrino-magnetic field driven kicks. Considering these three kick
mechanisms,
%that the kick timescales depend on different kick-driven
%modes.
Huang et al. (2003) found that the model of Dar \& Plaga
(1999) agrees well with the observations of GRBs. After
investigating the spectra and the light curve of SN2006aj associated
with an X-ray flash (GRB060218), Mazzali et al. (2006) found that
the progenitor of the burst is a star, whose initial mass was only
$\sim 20M_\odot$, expected to form only a residual neutron star
rather than a black hole when its core collapses.

In this work, we present a statistical model where the kick
velocity of a pulsar arises from the asymmetric explosion of a
mono-jet GRB.
Although the mechanism for the formation of one-side jet is based
neither on observational evidence nor on firm theoretical
evidence, the key point here as mentioned by Cen (1998) is to
couple a significant fraction of the total gravitational collapse
energy of the core to a very small amount of baryonic matter.
We suggest that SNe (to form pulsars) and GRBs are generally
associated, and try to know if the distribution of observed
pulsar's kicks and the modelled one from GRB luminosity are
statistically consistent.
After comparing the distribution of observed pulsar's kick
velocities with that from GRB energies, we find that these two
distributions may come from same parent population. In \S2, we
give the samples and equations for statistics. The statistical
results are presented in \S3. Conclusions and discussions are made
in \S4.

GRBs have been classified into long-soft and short-hard
categories. The former is currently supposed to be associated with
the death of massive stars, while the latter is suggested to be
related to the mergers of compact stars in elliptical/early-type
galaxies (Gehrels et al. 2005). We focus only on long-soft GRBs in
this paper.

%__________________________________________________________________

\section{Samples and equations}

From the ATNF pulsar
catalogue\footnote{http://www.atnf.csiro.au/research/pulsar/psrcat/},
121 isolated pulsars with known kick velocity are obtained, where
the archived velocity is the transverse velocity, i.e. the
projection of three-dimensional (3D) kick velocity on the celestial
sphere.
The GRB sample applied in this paper is currently the largest one
with known redshift\footnote{http://www.mpe.mpg.de/$\sim$jcg/},
$z$. It includes 98 GRBs, out of which 66 with known fluences
detected by BATSE (at 110-320 keV) or HETE II (30-400 keV) or
Swift (15-150 keV).

If a neutron star forms after the asymmetric explosion of a GRB,
other debris escape from one side with almost the speed of sight,
$c$, due to the huge energy released. According to the
conservation of momentum, the neutron star's momentum should be
$P_\mathrm{m}=E_{\gamma}/c$, with $E_{\gamma}$ the total energy of
GRBs. The kick velocity is then
$\upsilon=P_\mathrm{m}/M_{\mathrm{NS}}$. Here we adopt the mass of
neutron star, $M_{\mathrm{NS}}$, as the typical one of
$1.4M_\odot$, given by Stairs (2004) from the high-precision
pulsar timing observations.

% It is known that
The collimation-corrected total energy of GRBs from a conical jet
reads (Dado et al. 2006),
\begin{equation}
E_{\gamma}=\frac{1}{2}(1-\cos\theta)E_{\mathrm{iso}},\label{eqcoll}
\end{equation}
where $\theta$ is the opening angle of jet, and $E_{\mathrm{iso}}$
is the isotropic burst energy that is derived from observed
fluence and distance, or some models by assuming an isotropic
burst. Ghirlanda et al. (2004) suggested that the maximum opening
angle is about $\theta_{\mathrm{max}}=24^{\rm o}$. In this paper,
the opening angle for each GRB is generated randomly within
$\theta<\theta_{\mathrm{max}}$.

The method to calculate $E_{\mathrm{iso}}$, which applies to 66
GRBs with observed fluence $S$ and redshift $z$, is based on the
following relation,
\begin{equation}
 E_{\mathrm{iso}}=\frac{4\pi \kappa D_\mathrm{L}^2}{1+z}S,
\label{E`}
\end{equation}
where $D_\mathrm{L}$ is the luminosity distance of GRB, which, for
the sake of simplicity, is calculated by adopting
$\Omega_\mathrm{M}=0.3$, $\Omega_{\Lambda}=0.7$, and $H_0=71$ km
$\mathrm{s}^{-1}$ $\mathrm{Mpc}^{-1}$.
The factor $\kappa$ is applied  to convert the observed fluence at
observational energy band of an instrument (from $E_1$ to $E_2$,
in unit of keV) to that at a standard band in rest frame of GRB,
$(1-10^4)/(1+z)$ keV (Bloom et al. 2001), which reads,
\begin{equation}
\kappa=\frac{\int^{10^4/(1+z)}_{1/(1+z)}EN(E)dE}{\int^{E_2}_{E_1}EN(E)dE},
\label{k}
\end{equation}
where $E$ is photon energy, $N(E)$ is the band function defined by
Band et al. (1993) as follows
\begin{equation}
N(E)\propto\{^{E^{\alpha}e^{-E/E_0}\\
~~~~~~~~~~~~~~~~E\le(\alpha-\beta)E_0}_{[(\alpha-\beta)E_0]^{\alpha-\beta}
E^{\beta}e^{\alpha-\beta}\\
~~~E>(\alpha-\beta)E_0},
\end{equation}
where $\alpha$ and $\beta$ are spectral indices of GRBs. In our
calculation, the statistic mean spectral indices $\alpha \simeq
-1$, $\beta \simeq -2.2$ are substituted into $N(E)$ formula
(Preece et al. 2000). The peak photon energy, $E_0$, is adopted to
be $E_0 \simeq 200$ keV.

Substituting $E_{\rm iso}$ obtained with Eqs.(2-4) into
Eq.(\ref{eqcoll}) to calculate $E_\gamma$, one can figure out the
modelled kick velocity for each GRB. However, the velocity is
three dimensional ($\upsilon_{\mathrm{3D}}$). In order to compare
its distribution with that of the archived velocity of pulsars,
which is two dimensional velocity on celestial sphere, one needs
to use ``$\upsilon_{\mathrm{2D}}=\upsilon_{\mathrm{3D}}\cdot
\sin\phi$'' to do projection, where $\phi$ is the angle between
line of sight and ${\vec \upsilon}_{\mathrm{3D}}$. In our
calculation $\phi$ is obtained by generating a normalized random
number for $\sin\phi$ for each GRB, but not by generating a random
value for $\phi$ directly. That is because the direction of kick
velocity should be isotropic in space, the probability from $\phi$
to $\phi+{\rm d}\phi$ is proportional to $\sin \phi$ (note: it
could be easily obtained by considering the solid angle between
$\phi$ and $\phi+{\rm d}\phi$).

\section{Results}

With the modelled velocities, $v_{\rm 2D}$, obtained with above
equations, the distribution is plotted in Fig.1, as shown by
line with symbols. %grey histogram.
The histogram of ATNF-archived kick velocity of pulsars
(solid line) is also shown in the figure.
\begin{figure}
  \centering
    \includegraphics[width=10cm]{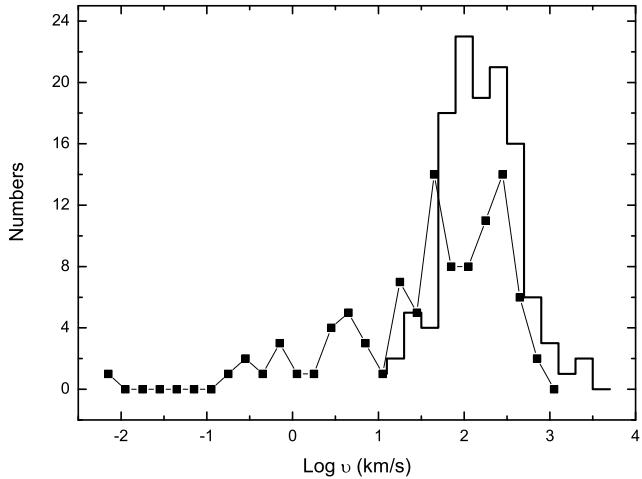}
    \caption{The distribution of pulsar's kick velocity and that derived from
    GRB model. %fluences.
    The solid step line is the observed kick distribution for the 121 pulsars in ATNF.
    The line with symbols is the modelled kick distribution %gray steps
    derived from the observed fluences and redshifts of GRB. %fluences  of GRBs with known redshifts.
    }
\end{figure}

The null hypothesis for two groups, i.e. the distributions of the
model above and observed velocities, is tested with
Kolmogorov-Smirnov (K-S) test. The maximum distance between their
cumulative probability functions is {$P_{\mathrm{KS}}$=0.36 on the
significant level $p=0.15$}. This indicates that the samples of
observed and modelled kick velocity could come from the same
parent population.

\section{Conclusions and discussions}

Based on the assumption that NSs are produced in asymmetric
explosion of GRBs associated with SNe and the conservation law of
momentum, we calculate the kick velocity of NSs from the
model %isotropic energy
of GRBs. Comparing the distribution of modelled kick velocity with
that of observed kick velocity of pulsars, it is found that two
distributions come from the same parent population. Therefore, we
conclude that the kick velocity of pulsars may come from the
asymmetric explosion of GRBs.
Our this work could be regarded as an observational test to the
idea proposed by Cen (1998) who suggested a unified scenario that
explains both pulsar kicks and cosmic GRBs.

In order to test the effect of the rotation period $P$, we
classified the pulsar sample into millisecond and normal pulsars,
and compare the distributions with modelled kick velocity by K-S
test. Designated $P=20$ ms as the critical period, there are 14
millisecond pulsars (hereafter ``$MSP14$'') and 107 normal pulsars
(``$NP107$''). The distributions of these two sub-samples are
compared with the GRB sample, i.e. ``$GRB66$'' (the modelled kick
velocity derived from the 66 GRBs with observed fluences and
redshifts), via K-S test. The results are listed in Table 1. In
the bracket is the significant level $p$ for the corresponding
maximum distance $P_{\mathrm{KS}}$ between the cumulative
probability functions.

The effect of the characteristic age $\tau$ is also tested.
Assigning the characteristic age $\tau_0=4\times 10^6$ yrs, we
find 44 pulsars with $\tau <\tau_0$ (sub-sample ``$Young44$'') and
73 with $\tau > \tau_0$ (sub-sample ``$old73$''). Note that there
are 4 pulsars without detected ages. The results of K-S test are
also presented in Table 1.

\begin{table}[]
  \caption[]{The K-S test results for the kick velocity of pulsar
   sub-classes and GRB sample. Numbers of $P_{\mathrm{KS}}(p)$ are listed,
   with $P_{\mathrm{KS}}$ the maximum distance between the cumulative
   probability functions and $p$ the significant level.}
  \label{Tab:publ-works}
  \begin{center}\begin{tabular}{cccccc}
  \hline\noalign{\smallskip}
Sample &  $MSP14$  & $NP107$   & $Young44$  &   $old73$  \\
  \hline\noalign{\smallskip}
% $GRB66$ & 0.33 (0.43) & 0.38 (0.15) & 0.13 (0.89)   & 0.24 (0.69)  \\
$GRB66$ & 0.30 (0.52) & 0.33 (0.25) & 0.18 (0.90)   & 0.17 (0.92)
\\
  \noalign{\smallskip}\hline
  \end{tabular}\end{center}
\end{table}

From Table 1, all the values of significant level for
$P_\mathrm{KS}$ are larger %enough
than $0.05$ to indicate that two sub-samples in any pair come from
the same parent population. It implies that the consistence of
modelled and observed kick velocity distributions
%the distribution of pulsar kick velocities
may be intrinsic and does not change with pulsar's periods or
ages.

In summary, a primary statistical test to the consistence of the
kick velocity distributions from the model of SN-related GRB and
from the observations of pulsars
%SN-GRB association
is done via K-S test. %comparing pulsar's kicks and GRB fluences.
Advanced research to check the idea that asymmetric fireballs
result in kicks is needed as more related observational data would
be possible in the future.
Statistically, we find that the distribution of observed
%observation of
pulsar's kicks may
be consistent with that deduced from the model of %the detected energy of
GRBs under the assumption that pulsar's kicks arise from the
one-side explosion of SN-related GRB.
%asymmetry of GRB fireballs.
Comprehensive understanding on this statistics in
theory is still not certain and is very necessary.

\begin{acknowledgements}

The authors thank helpful discussion with the members in the
pulsar group of Peking University. The helpful comments and
suggestions from an anonymous referee are sincerely acknowledged.
This work is supported by NSFC (10573002, 10778611) and by the Key
Grant Project of Chinese Ministry of Education (305001).

\end{acknowledgements}

\end{document}